%% file: grqc-2sphere.tex
\documentclass[aps,prd,showpacs,preprint]{revtex4}
\usepackage[german,british]{babel}
\usepackage{amsmath}
\usepackage{amssymb}

\DeclareFontFamily{OT1}{rsfs}{}
\DeclareFontShape{OT1}{rsfs}{m}{n}{<-7> rsfs5 <7-10> rsfs7 <10->rsfs10}{}
\DeclareMathAlphabet{\mycal}{OT1}{rsfs}{m}{n}

\usepackage{graphicx}
\usepackage{color}

\begin{document}

\title{A new two-sphere singularity in general relativity}

\author{Christian G.~B\"ohmer}
\email{christian.boehmer@port.ac.uk}
\affiliation{Institute of Cosmology \& Gravitation,
             University of Portsmouth, Portsmouth PO1 2EG, UK}

\author{Francisco S.~N.~Lobo}
\email{francisco.lobo@port.ac.uk}
\affiliation{Institute of Cosmology \& Gravitation,
             University of Portsmouth, Portsmouth PO1 2EG, UK}

\date{\today}

\begin{abstract}

The Florides solution, proposed as an alternative to the interior Schwarzschild
solution, represents a static and spherically symmetric geometry with vanishing
radial stresses. It is regular at the center, and is matched to an exterior
Schwarzschild solution. The specific case of a constant energy density has been
interpreted as the field inside an Einstein cluster. In this work, we are interested
in analyzing the geometry throughout the permitted range of the radial coordinate
without matching it to the Schwarzschild exterior spacetime at some constant radius
hypersurface. We find an interesting picture, namely, the solution represents a
three-sphere, whose equatorial two-sphere is singular, in the sense that the
curvature invariants and the tangential pressure diverge. As far as we know, such
singularities have not been discussed before. In the presence of a large negative
cosmological constant (anti-de Sitter) the singularity is removed.

\end{abstract}

\pacs{04.20.-q, 04.20.Dw, 04.40.Nr}
\maketitle

\section{Introduction}

The construction of theoretical models describing relativistic
stars and the phenomenon of gravitational collapse is a
fundamental issue in relativistic astrophysics. Pioneering work
was done by Schwarzschild~\cite{Schw}, who analyzed solutions
describing a star of uniform energy density; Tolman provided
explicit solutions of static fluid spheres~\cite{Tolman};
Oppenheimer and Volkoff~\cite{OppVolk}, by considering specific
Tolman solutions, analyzed the gravitational equilibrium of
stellar structures; Oppenheimer and Snyder~\cite{OppSnyd},
provided the first insights to gravitational collapse into a black
hole; Buchdahl~\cite{Buchdahl} and Bondi~\cite{Bondi} also
generalized the interior constant energy density solutions to more
general static fluid spheres in the form of inequalities involving
the energy density, central pressure and the location of the
boundary matching surface. These authors, amongst others, lay down
the foundations of the general relativistic theory of stellar
structures (see Ref.~\cite{Gravitation} for an extensive review).

In the 1970's, Florides in an attempt to understand, within the
framework of general relativity, why a spherically symmetric
distribution of pressure-less dust at rest cannot maintain itself
in equilibrium, discovered a new interior uniform density
(Schwarzschild-like) solution. The latter solution is static,
spherically symmetric, regular throughout the interior, and is
matched to an exterior Schwarzschild spacetime~\cite{Florides}. It
is interesting to note that the radial pressure is identically
zero and the tangential pressure is positive and an increasing
function of the radial coordinate. Now, at a first glance the
absence of a radial pressure may cast doubts upon the physical
significance of the Florides solution, as one is accustomed to
thinking that it is precisely this radial pressure that maintains
a system in static equilibrium. However, it was found to have a
rather elegant physical interpretation, namely, the Florides
interior solution describes the interior field of an Einstein
cluster~\cite{Einstein} (The Florides solution was further
analyzed in Ref.~\cite{Kofinti}). Recall that the Einstein cluster
describes a static and spherically symmetric gravitational field
of a large number of particles moving in randomly oriented
concentric circular orbits under their own gravitational field.

Whilst analyzing the Florides solution within itself, and
considering the whole permitted range of the radial coordinate,
without the respective matching to an exterior Schwarzschild
spacetime at a junction interface, we came across an extremely
interesting feature, namely, that the solution in fact represents
a three-sphere, possessing a singular equatorial two-sphere, in
the sense that the curvature invariants and the tangential
pressure diverge. This interesting aspect of the geometry
motivated a more careful analysis of the Florides solution, as
these two-sphere singularities have not been investigated before,
to the best of our knowledge. However, it is interesting to note
that in Ref.~\cite{Ellis}, by dropping the assumption of
homogeneity on a cosmological scale, the authors considered a
static and spherically symmetric model, containing a singularity
which continually interacts with the Universe. It was suggested
that the singularity can be interpreted as a sphere surrounding a
regular central region. But, it is important to emphasize that the
latter singularity is fundamentally different in nature to the
two-sphere singularity analyzed in this work, as shall be
discussed below in more detail.

We also stress that spacetime singularities have played a
fundamental role in conceptual discussions of general relativity,
and a key aspect of singularities in general relativity is whether
they are a disaster for the theory, as they imply the breakdown of
predictability. One may mention several attitudes that are
widespread in the literature~\cite{Earman}, namely, that
singularities are mere artifacts of unrealistic and idealized
models; general relativity entails singularities, but fails to
accurately describe nature; and the existence of singularities may
be viewed as a source to probe the limitations of general
relativity, and from which one may derive a valuable understanding
of cosmology~\cite{Misner}. We adopt the latter viewpoint
throughout this work, attempting to understand the nature of the
two-sphere singularity present in the Florides solution.

This paper is outlined in the following manner: In
Section~\ref{Sec:2}, we deduce the general radial pressure-less
solution, and further consider the specific case of constant
energy density, with and without a cosmological constant, and
provide the geometrical interpretation of the singular two-sphere.
In Section~\ref{Sec:3}, we analyze specific characteristics of the
geometry, such as the conserved quantities and geodesic motion.
Finally, in Section~\ref{Sec:4}, we conclude.

\section{Interior constant density solutions re-analyzed}
\label{Sec:2}

\subsection{General radial pressure-less solution}

Consider a static and spherically symmetric spacetime, given in
curvature coordinates, by the following line element
\begin{equation}
      ds^2=-e^{2\alpha(r)}\,dt^2 +e^{2\beta(r)}\,dr^2+r^2\,d\Omega^2\,.
      \label{generalmetric}
\end{equation}
where $d\Omega^2 = d\theta^2 + \sin^2\theta \,d\phi^2$. As we are
interested in analyzing solutions with a vanishing radial
pressure, i..e, $p_r(r)=0$, in the spirit of Ref.~\cite{Florides},
the anisotropic stress energy tensor is given by
\begin{equation}
T_{\mu\nu}=\rho\,U_\mu \, U_\nu+p_\perp\, g^\perp_{\mu\nu} \,,
\end{equation}
where $g^\perp_{\mu\nu}$ is the projection of the metric along the
transverse spatial direction, i.e., orthogonal to the radial
direction. It is defined as
$g^\perp_{\mu\nu}=g_{\mu\nu}+U_\mu\,U_\nu-\chi_\mu\,\chi_\nu$,
where $U^\mu$ is the four-velocity, and $\chi^\mu$ is the unit
spacelike vector in the radial direction, i.e.,
$\chi^\mu=e^{-\beta}\,\delta^\mu{}_r$. Note that
$g^\perp_{\mu\nu}U^\nu=0$, $g^\perp_{\mu\nu}\chi^\nu=0$ and
$U_{\mu}\,\chi^\mu=0$. $\rho(r)$ is the energy density, and
$p_\perp(r)$ is the transverse pressure measured in the orthogonal
direction to $\chi^\mu$.

Using the Einstein field equation, $G_{\mu\nu}=8\pi \,T_{\mu\nu}$
(with $c=G=1$), the stress energy tensor components are given by
\begin{eqnarray}
8\pi \rho(r) &=& \frac{e^{-2\beta}}{r^2}
\,\left(2\beta'r+e^{2\beta}-1 \right)
\label{rho}\,,\\
8\pi p_r(r) &=& \frac{e^{-2\beta}}{r^2}
\,\left(2\alpha'r-e^{2\beta}+1 \right)
\label{pr}=0\,,\\
8\pi p_\perp(r) &=& \frac{e^{-2\beta}}{r}
\,\left[-\beta'+\alpha'+r\alpha''+r(\alpha')^2-r\alpha' \beta'\right]
\label{pt}\,,
\end{eqnarray}
where the prime denotes a derivative with respect to the radial
coordinate $r$.

Integration of Eq.~(\ref{rho}) yields the following relationship
\begin{equation}
e^{-2\beta(r)}=1-\frac{2m(r)}{r}\,,    \qquad  {\rm with} \qquad
m(r)=4\pi\int_0^r\rho(\bar{r})\bar{r}^2\,d\bar{r} \,,
    \label{beta}
\end{equation}
where the integration constant has been evaluated by considering
$\beta(0)=0$. The function $m(r)$ is the quasi-local mass, and is
denoted as the mass function. Substituting Eq.~(\ref{beta}) in Eq.
(\ref{pr}), we have
\begin{equation}
2\alpha(r)= \int_a^r
\frac{2m(\bar{r})}{\bar{r}^2(1-2m(\bar{r})/\bar{r})} \,d\bar{r} +C\,,
\label{alpha}
\end{equation}
where the constant of integration may be determined by matching
this interior solution to a Schwarzschild exterior solution at a
junction interface, $a$. Thus, the constant is given by
$C=\ln(1-2M/a)$, where $M$ is the object's mass, with $a>2M$. With
these relationships the tangential pressure is given by
\begin{equation}
p_\perp(r)= \frac{m(r)\rho(r)}{2r(1-2m(r)/r)}  \,.
\end{equation}
The latter relationship may also be obtained from the conservation
of the stress energy tensor, $\nabla_{\nu}T^{\mu\nu}=0$, which
provides the anisotropic form of the Tolman-Oppenheimer-Volkoff (TOV)
equation. The metric finally assumes the form
\begin{equation}
ds^2=-\left(1-\frac{2M}{a}\right)\,\left[\exp\int_a^r
\frac{2m(\bar{r})\,d\bar{r}}{\bar{r}^2(1-2m(\bar{r})/\bar{r})}
\right]\,dt^2 +\frac{dr^2}{1-2m(r)/r}+r^2\,d\Omega^2 \,.
\label{generalmetric2}
\end{equation}

\subsection{Constant energy density}

Considering a constant energy density~\cite{Florides}, the mass
function, given by Eq.~(\ref{beta}), and Eq.~(\ref{alpha}) are
readily integrated. This provides metric (\ref{generalmetric2}) in
the following simplified form
\begin{align}
      ds^2 = -\frac{\Bigl(1-\frac{8\pi}{3} \rho_0 a^2 \Bigr)^{3/2}}
      {\Bigl(1-\frac{8\pi}{3} \rho_0 r^2 \Bigr)^{1/2} } dt^2 +
      \frac{dr^2}{1-\frac{8\pi}{3} \rho_0 r^2} + r^2 d\Omega^2 \,,
      \label{metric1}
\end{align}
and corresponds to the following stress energy tensor
\begin{align}
      \rho(r) = \rho_0\,,\qquad p_r(r) = 0\,, \qquad
      p_\perp(r) = \frac{2\pi \rho_0^2 r^2}
      {3 (1-\frac{8\pi}{3} \rho_0 r^2)} \,,
      \label{stresses}
\end{align}
where $a$ is the matching surface, and the time was scaled to
match with the Schwarzschild exterior. We immediately verify some
problems with the above stress energy tensor components in
comparison with the usual Schwarzschild interior solution. In the
case of an isotropic constant density perfect fluid, the field
equations yield a closed system of equations that can be solved
uniquely for any given central pressure by means of the
Tolman-Oppenheimer-Volkoff equation. The resulting pressure
function turns out to be monotonically decreasing and its
vanishing uniquely defines the boundary of the stellar object at
which the Schwarzschild exterior metric can be matched.

Looking at the above tangential pressure component, given by
Eq.~(\ref{stresses}), we realize that the pressure is
monotonically increasing, and in fact diverges as $r\rightarrow
1/\sqrt{8\pi\rho_0/3}$. Hence, in contrast to the Schwarzschild
interior solution, there is no `preferred' vanishing pressure
surface that is implied by the field equations. Thus one may match
this interior solution at any value $2M<a<1/\sqrt{8\pi\rho_0/3}$
to an exterior Schwarzschild spacetime and thereby avoiding the
discussion of the singularity at $r=1/\sqrt{8\pi\rho_0/3}$. This
feature of the solution motivates the geometrical analysis of the
global spacetime described by metric~(\ref{metric1}). Florides and
later authors certainly noticed the divergent tangential pressure
but presumably avoided its discussion by requiring that the
Schwarzschild metric is matched at some smaller radius.

The chosen coordinate system for metric~(\ref{metric1}) is defined
for radii that satisfy $r<R$. It is however easy to introduce a
coordinate system that yields a much better geometrical
understanding. Therefore, let us introduce a third angle $\alpha$
defined by
\begin{align}
      r = R \sin \alpha\,, \qquad  {\rm with} \qquad
      1/R = \sqrt{\frac{8\pi\rho_0}{3}}\,,
\end{align}
for which the metric~(\ref{metric1}) becomes
\begin{align}
      ds^2 = -\frac{\cos^3\negmedspace\alpha_b}{\cos\alpha} dt^2 +
      R^2 d\alpha^2 + R^2 \sin^2\negmedspace\alpha d\Omega^2 \,,
      \label{eq:m3}
\end{align}
where $a=R\sin\alpha_b$. The metric coefficient $g_{tt}$ can be
rescaled to have $g_{tt}(\alpha=0)=1$, which leads to the form of
the metric that will be used henceforth
\begin{align}
      ds^2 = -\frac{dt^2}{\cos\alpha} +
      R^2 d\alpha^2 + R^2 \sin^2\negmedspace\alpha d\Omega^2 \,.
      \label{eq:m4}
\end{align}
In performing a more careful analysis of the Florides solution we
come across an extremely interesting aspect, namely, that the solution
in fact represents a three-sphere, possessing a singular equatorial
two-sphere. Before analyzing the geometry of the spacetime in more
detail, we shall briefly consider the inclusion of a cosmological constant.

\subsection{Presence of a cosmological constant}

Chongming {\it et al}~\cite{Chong} extended the original work of
Florides by taking into account the cosmological constant. This
generalization yields the following metric and stress energy
tensor, given by
\begin{align}
      ds^2 = -\frac{\Bigl(1-\frac{8\pi}{3}\rho_0 a^2-\frac{\Lambda}{3}a^2
      \Bigr)^{3/2(1+\Lambda/8\pi\rho_0)}}
      {\Bigl(1-\frac{8\pi}{3}\rho_0 r^2-\frac{\Lambda}{3}r^2
      \Bigr)^{(1-\Lambda/4\pi\rho_0)/2(1+\Lambda/8\pi\rho_0)}} dt^2 +
      \frac{dr^2}{1-\frac{8\pi}{3}\rho_0 r^2-\frac{\Lambda}{3}r^2} + r^2 d\Omega^2 \,,
      \label{eq:x1}
\end{align}
and
\begin{align}
      \rho(r) = \rho_0\,, \qquad p_r (r) = 0\,, \qquad
      p_\perp(r) = \frac{2\pi \rho_0^2 r^2}
      {3 \Bigl(1-\frac{8\pi \rho_0}{3}r^2-\frac{\Lambda}{3}r^2\Bigr)}
      \Bigl(1-\frac{\Lambda}{4\pi\rho_0}\Bigr) \,,
      \label{eq:x2}
\end{align}
respectively. It is interesting to note that the spatial geometry
now depends on the cosmological constant, see
e.g.~\cite{Weyl:1919,Stuchlik:2000,Boehmer:2002gg,Boehmer:2003uz,Kozyrev:2004kg}.
Let us introduce a new parameter $k$ defined by
\begin{align}
      k = \frac{8\pi \rho_0}{3}+\frac{\Lambda}{3} \,.
      \label{eq:x3}
\end{align}
For $k>0$ the spatial geometry corresponds to a space of constant
positive curvature, i.e., a sphere; for $k=0$ the geometry is
Euclidean; and for $k<0$ the spatial geometry has constant
negative curvature and is therefore hyperbolic. It is then also
useful to introduce, similar to the third angle in the spherical
case, adapted coordinates for the hyperbolic case. Hence, define
\begin{align}
      r = \frac{1}{\sqrt{|k|}}\sinh\alpha \,,
     \label{eq:x4}
\end{align}
so that for the specific case of $k<0$, metric~(\ref{eq:x1}) takes
the form
\begin{align}
      ds^2_{k<0} = -\frac{\cosh^3\negmedspace\alpha_b^{1/(1+\Lambda/8\pi\rho_0)}}
      {\cosh\alpha^{(1-\Lambda/4\pi\rho_0)/(1+\Lambda/8\pi\rho_0)}} dt^2 +
      \frac{1}{|k|}(d\alpha^2 + \sinh\negmedspace^2\alpha d\Omega^2) \,.
      \label{eq:x5}
\end{align}
Similarly, the stress energy tensor components simplify in the
new coordinates to the following relationships
\begin{align}
      \rho(r) = \rho_0\,, \qquad p_r(r) = 0\,, \qquad
      p_\perp(r)  = \frac{2\pi \rho_0^2}{|k|}
      \Bigl(1-\frac{\Lambda}{4\pi\rho_0}\Bigr) \tanh^2\negmedspace\alpha \,.
      \label{eq:x6}
\end{align}
In contrast to the spherical case, the tangential pressure does
not diverge in the hyperbolic case since
$\lim_{\alpha\rightarrow\infty}\tanh\alpha = 1$. Furthermore, the
center $\alpha=0$ is regular (even flat) and therfore this
hyperbolic spacetime is a globally regular spacetime, completely
filled with an anisotropic perfect fluid, having constant energy
density and vanishing radial pressure. Moreover, the
Schwarzschild-anti de Sitter spacetime can be matched at any
$\alpha = {\rm constant}$ hypersurface so that the induced metric
and the extrinsic curvature with respect to the matching surface
are both continuous.

For the specific Euclidean case, where $k=0$, one has to be
careful by appropriately taking the limit in metric~(\ref{eq:x1})
because of the exponent. For consistency of the notation, let us
rename $r$ by $\alpha$, so that the metric reads
\begin{align}
      ds^2_{k=0} = -\exp\Bigl(4\pi\rho_0 (\alpha^2-\alpha_b^2) \Bigr) dt^2 +
      (d\alpha^2 + \alpha^2 d\Omega^2) \,,
      \label{eq:x7}
\end{align}
having the stress energy tensor components
\begin{align}
      \rho(\alpha) = \rho_0\,, \qquad p_r(\alpha) = 0\,, \qquad
      p_\perp(\alpha)  = 2\pi \rho_0^2 \alpha^2\,.
      \label{eq:x7a}
\end{align}

It should also be noted that the stress energy
tensor~(\ref{eq:x2}) implies that for $\Lambda=4\pi\rho_0$, the
tangential pressure also vanishes. In this case the stress energy
tensor reduces to pressure-less dust, $\rho(r)=\rho_0$. Since the
$k$ is also positive, the spatial geometry is spherical and hence
this spacetime is the original Einstein static universe
(pressure-less) that was suggested by Einstein in 1917. However,
the spherical case with non-vanishing tangential pressure does not
allow the construction of an anisotropic Einstein static universe
with vanishing radial pressure. It should be noted that the
Einstein static universe can be generalized to have non-constant
pressure with two regular centers,
see~\cite{Ibrahim:1976,Boehmer:2002gg,Boehmer:2003uz,Boehmer:2004nu}, and also
in spacetimes with torsion an analog Einstein universe can be
constructed, containing a constant radially symmetric torsion
field~\cite{Boehmer:2003iv}. It seems therefore that the
anisotropic Einstein static universe is much more difficult to
construct and it may possibly require a non-constant energy
density. A similar question has not been answered yet (as far as
we know), namely if a charged Einstein universe can in principle
be constructed that also is globally regular.

Another interesting feature of an anisotropic matter distribution
is their recent appearance in a rather different context of
gravastars and dark energy stars (see e.g. Ref.~\cite{Lobo:2005uf}
and references therein). For a constant energy density, the metric
reported in~\cite{Lobo:2005uf} takes the form
\begin{align}
      ds^2 = -(1-2Ar^2)^{-(1+3w)/2}dt^2 + \frac{dr^2}{1-2Ar^2} + r^2 d\Omega^2 \,,
      \label{eq:x8}
\end{align}
where, as above, one can easily introduce a new coordinate (third
angle) by $r=(1/\sqrt{2A})\sin\alpha$. One should note that $w$,
the dark energy equation of state parameter, and $A$ can both be
chosen so that this metric agrees with either the Florides
metric~(\ref{metric1}), where simply $w=0$, or its generalization
due to the presence of $\Lambda$, Eq.~(\ref{eq:x1}). However, it
is important to emphasize that to be a gravastar (or a dark energy
star) solution a fundamental ingredient is a repulsive interior
spacetime. This differs from the Florides solution, as in the
latter the interior geometry is attractive.

\subsection{Two-sphere singularity}

We now turn to the analysis of the nature of the singularity in
metric (\ref{eq:m4}), for the spherically symmetric case with
$\Lambda=0$, although the main results that follow are unchanged
for $k>0$. For that, let us compute the non-vanishing Riemann
tensor components
\begin{align}
      R_{\alpha t}{}^{\alpha t} =
      \frac{\cos^2\negmedspace\alpha-3}{4R^2\cos^2\negmedspace\alpha}\,,\\
      R_{\alpha\theta}{}^{\alpha\theta} = R_{\alpha\phi}{}^{\alpha\phi}
      = R_{\theta\phi}{}^{\theta\phi} = \frac{1}{R^2} \,,\\
      R_{\theta t}{}^{\theta t} = R_{\phi t}{}^{\phi t} = -\frac{1}{2R^2}\,,
\end{align}
and Weyl tensor components
\begin{align}
      C_{\alpha\theta}{}^{\alpha\theta} = C_{\alpha\phi}{}^{\alpha\phi}
      = C_{\theta t}{}^{\theta t} = C_{\phi t}{}^{\phi t}
      = \frac{\tan^2\negmedspace\alpha}{8 R^2}\,,\\
      C_{\alpha t}{}^{\alpha t} = C_{\theta\phi}{}^{\theta\phi}
      = -\frac{\tan^2\negmedspace\alpha}{4R^2}\,,
\end{align}
respectively. Since these Weyl tensor components are
non-vanishing, we note crucial geometrical differences between the
interior Schwarzschild solution and the Florides solution. It is
well-known that the Schwarzschild interior solution is conformally
flat~\cite{Weyl:1919,Stephani}, irrespective of the cosmological
constant~\cite{Boehmer:2002gg,Boehmer:2003uz}.

These yield the squared Riemann and Weyl tensors,
and we also note the square of the Ricci tensor
\begin{align}
      {\rm RiemSq} &= \frac{3(68\cos(2\alpha)+19\cos(4\alpha)+73)}
      {32 R^4 \cos^4\negmedspace\alpha}\,,\\
      {\rm WeylSq} &= \frac{3\tan^4\negmedspace\alpha}{4 R^4}\,,\\
      {\rm RicciSq} &= \frac{9(28\cos(2\alpha)+9(\cos(4\alpha )+3))}
      {64 R^4 \cos^4\negmedspace\alpha}\,,
\end{align}
respectively. All three geometrical invariants diverge near
$\alpha=\pi/2$ in a similar way, by which we mean
\begin{align}
      \lim_{\alpha\rightarrow\pi/2}\frac{\rm WeylSq}{\rm RicciSq} &= \frac{2}{3}\,,\\
      \lim_{\alpha\rightarrow\pi/2}\frac{\rm RiemSq}{\rm RicciSq} &= \frac{2}{3}\,,\\
      \lim_{\alpha\rightarrow\pi/2}\frac{\rm WeylSq}{\rm RiemSq} &= \frac{1}{2}\,,
\end{align}
namely, the Weyl tensor is not dominated by the Ricci
tensor~\cite{Goode:1985ab} and the singularity does not correspond
to an isotropic singularity. The Ricci scalar is given by
\begin{align}
      g^{\mu\nu}R_{\mu\nu}=\frac{3(2-3\sin^2\alpha)}{2R^2\cos^2\alpha}\,,
\end{align}
which also diverges at $\alpha=\pi/2$.

\begin{figure}[!h]
\noindent
\begin{minipage}[h]{.96\linewidth}
\begin{center}
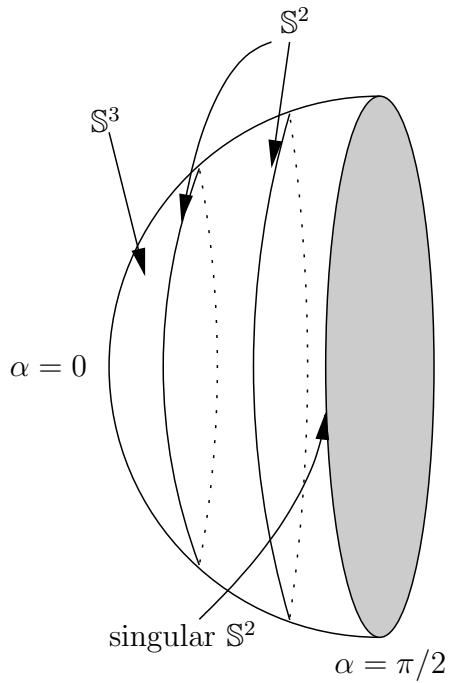
\end{center}
\end{minipage}\\[1ex]
\caption{This figure represents the spatial three-sphere $\mathbb{S}^3$.
         Vertical cuts through the three-sphere define the two-spheres
         $\mathbb{S}^2$ of the spherically symmetric spacetime. The
         equatorial cut through $\mathbb{S}^3$ at $\alpha=\pi/2$ defines
         the singular two-sphere $\mathbb{S}^2$.}
\label{figa}
\end{figure}

It is interesting to note that such a singularity is not
point-like. It describes a singular two-sphere, but the spacetime
is well defined for $\alpha\in[0,\pi/2)$. Since the spatial part
of this spacetime is a three-sphere we find the following
geometrical picture: a three-sphere whose equatorial two-sphere is
singular in the sense that the above invariants and the tangential
pressure diverge. However, the radial pressure (identically zero)
and the energy density are both finite at the singularity.
Fig.~\ref{figa} represents the spatial three-sphere $\mathbb{S}^3$.

\begin{figure}[!h]
\noindent
\begin{minipage}[h]{.96\linewidth}
\begin{center}
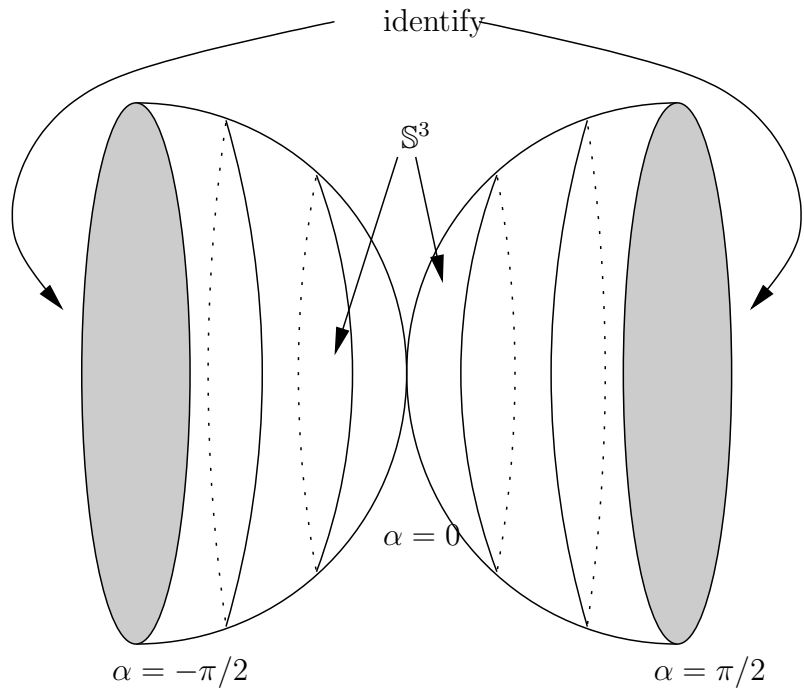
\end{center}
\end{minipage}\\[1ex]
\caption{This figure represents the two half spatial three-spheres $\mathbb{S}^3$.
         At $\alpha=0$ they have a common point and we identify both singular
         two-spheres.}
\label{figb}
\end{figure}

It should be noted however, that the metric~(\ref{eq:m4}) is actually well
defined for $\alpha=(-\pi/2,\pi/2)$. Therefore, we could in principle draw
a second copy of the three-sphere so that the two half three-spheres are
joined at $\alpha=0$ rather than at the equator. However, we can still identify
both singular two-spheres and would obtain something like Fig.~\ref{figb}.

The Florides solution can be interpreted as the interior of an
Einstein cluster, therefore the singularity could also be
interpreted from that point of view. We have a large number of
particles that move in oriented circular orbits. Their individual
velocities and angular momenta~\cite{Lake:2006pp} are related to
the tangential pressure and therefore the singular two-sphere
corresponds to the surface where the particles all move with the
speed of light. Obviously, such a surface has singular properties
and it is expected that the proper time of the geodesics is zero,
which is shown explicitly in the next Section. Furthermore our
analysis seems to be important also in the context of rotating
magnetized stellar objects. There exists a similar phenomena with
respect to the rotating magnetic field lines. At large distances
from the surface of the stellar object these line would rotate
with the speed of light. A more detailed study comparing the
geometry of the rotating magnetic field lines with the
analytically simple Florides solution might shed some new light to
the theory of magnetized stellar objects.

As referred to in the Introduction, in Ref.~\cite{Ellis}, the
authors proposed a static and spherically symmetric model of the
Universe, containing a singularity which can be viewed as a sphere
surrounding a central region $C$, at $r=0$. The solution possesses
two centers, one at $r=0$ and the second at $r=R$, as the surface
area of the two-spheres of symmetry tends to zero at both centers.
All past radial null geodesics intersect the singularity, as do
all spacelike radial geodesics. Thus, the singularity can be
interpreted as surrounding the central region $C$, and lies within
a finite distance from $C$, so that the Universe is spatially
finite and bounded. Note that traversing a radial null geodesic
from the singularity, one reaches the central region after a
distance $R$, and the singularity is attained again after
traveling a total distance of $2R$. The spacetime can be thought
of as being spherically symmetric about both $C$ and the singular
center (we refer the reader to Ref.~\cite{Ellis} for details).
Note that the nature of this singularity is fundamentally
different to the two-sphere singularity in the Florides
solution. First, in Ref.~\cite{Ellis} the stress energy tensor
components tend to zero as the singularity is attained at
$r\rightarrow R$, while in the Florides solution the
tangential pressure diverges. Second, the geometric structure is
different, as the surface area of the Florides two-sphere is
monotonically increasing, contrary to the case analyzed in
Ref.~\cite{Ellis}, where the surface area is zero at $r=0$ and at
the singularity $r=R$. Thirdly, the cases analyzed in
Ref.~\cite{Ellis} impose $g_{tt}\rightarrow 0$ at $r\rightarrow R$
(although the case $g_{tt}\rightarrow \infty$ is briefly hinted
at, it is not analyzed), while in the Florides solution we have
$g_{tt}\rightarrow \infty$ as $r\rightarrow R$.

\section{The geodesic structure of these solutions}
\label{Sec:3}

\subsection{Conserved quantities}

Throughout this section, we shall consider metric (\ref{eq:m4}).
Consider the following Lagrangian
\begin{equation}
{\cal L}(x^\mu,\dot{x}^\mu)=\frac{1}{2}g_{\mu\nu}\dot{x}^\mu
\dot{x}^\nu \,.
\end{equation}
If the metric tensor does not depend on a determined coordinate,
$x^\mu$, through the Euler-Lagrange equations one obtains that the
quantity
\begin{equation}
      \pi_\mu=\frac{\partial {\cal L}}{\partial
      \dot{x}^\mu}=g_{\mu\nu}\,\dot{x}^\nu  \,,
      \label{conserv-quant}
\end{equation}
is constant along any geodesic. Applied to line element~(\ref{eq:m4}), one
verifies that the metric tensor components are independent of the coordinates
$t$ and $\phi$, so that the conserved quantities are given by
\begin{eqnarray}
\pi_{\phi}&=&g_{\phi\phi}\,\dot{\phi}=R^2\,\sin^2\alpha\;\dot{\phi}=L  \,,  \label{P}  \\
\pi_t&=&g_{tt}\,\dot{t}= -\frac{\dot{t}}{\cos\alpha}=-{\cal E}
\label{E}\,.
\end{eqnarray}
${\cal E}$ and $L$ may be interpreted as the energy and angular
momentum per unit mass. Without a loss of generality we may
consider the equatorial plane with $\theta=\pi/2$.

The line element (\ref{eq:m4}) may be rewritten in terms of the
constants defined above, for the particular case of
$\theta=\pi/2$, in the following manner
\begin{equation}
R^2\dot{\alpha}^2={\cal E}^2\,\cos\alpha+ \left(2{\cal
L}-\frac{L^2}{R^2\,\sin^2\alpha}\right) \,, \label{line2}
\end{equation}
where ${\cal L}=0$ is defined for null geodesics, and ${\cal
L}=-1/2$ for timelike geodesics.

The values of ${\cal E}$ and $L$ are determined by the initial
conditions of the movement. For instance, consider a fixed
observer along a point on the geodesic. The velocity of a test
geodesic particle (see Ref.~\cite{Doran} for details), as measured
by the observer is given by
\begin{equation}
V^2=\frac{\cos\alpha}{\dot{t}^2}\,(R^2
\dot{\alpha}^2+R^2\,\sin^2\alpha\,\dot{\phi}^2 ) \,,
\end{equation}
and substituting Eq.~(\ref{P})-(\ref{line2}), we have
\begin{equation}
{\cal E}^2=\frac{1}{(1-V^2)\cos\alpha} \,,
\end{equation}
for timelike geodesics, ${\cal L}=-1/2$. If a body initiates its
movement at $\alpha=0$ ($r=0$) with $v=0$, then ${\cal E}=1$. Note
that at $\alpha=\pi/2$ ($r=R$), we have ${\cal E}=\infty$. Indeed,
the range of ${\cal E}$ is precisely $1<{\cal E}^2<\infty$, as
shall also be shown below.

\subsection{Geodesics}

\subsubsection{Null geodesics}

Consider null geodesics along the $\alpha-$direction
($r-$direction), i.e., with $d\theta=d\phi=0$, so that $dt=\pm
\sqrt{\cos\alpha}\,R\,d\alpha$. Integrating the latter provides
the following solution
\begin{equation}
t=\pm 2R \;{\rm E}(\alpha/2,2) +C_1 \,,
\end{equation}
where ${\rm E}(\alpha,m)$ is the elliptic function of the second
kind, defined as
\begin{equation}
{\rm E}(\alpha,m)=\int_0^\alpha\,\sqrt{1 - m \sin\negmedspace^2\varphi}\,d\varphi \,.
\end{equation}

\begin{figure}[ht]
\centering
  \includegraphics[width=0.48\linewidth]{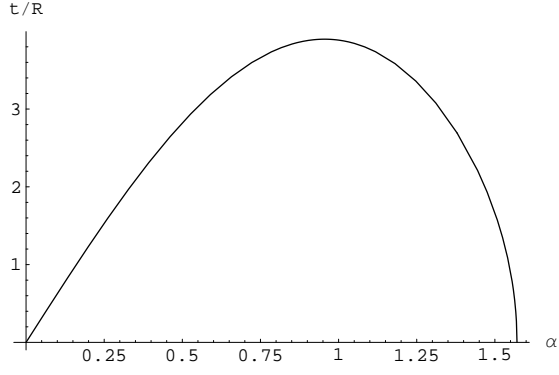}
  \caption{Consider the plot of circular null geodesics
  $t/R=2\pi \sin\alpha \sqrt{\cos\alpha}$. Note that the time
  coordinate attains a maximum at $\alpha=\arcsin\sqrt{2/3}$, and
  tends to zero as $\alpha\rightarrow \pi/2$ ($r\rightarrow R$).}
  \label{Fig:circ}
\end{figure}

Consider the specific case of circular orbits, i.e.,
$\theta=\pi/2$ and $d\alpha=0$, so that the line element reduces
to
\begin{equation}
ds^2=-\frac{dt^2}{\cos\alpha}+R^2\sin^2\alpha\,d\phi^2 \,.
\end{equation}
The null geodesic, $ds^2=0$, provides $dt/d\phi=\pm R\sin\alpha
\sqrt{\cos\alpha}$, which has the following solution
\begin{equation}
t=2\pi R\sin\alpha \sqrt{\cos\alpha}  \,.
\end{equation}
Note an interesting feature of this spacetime, namely, the time
coordinate tends to zero as $\alpha\rightarrow \pi/2$ ($r
\rightarrow R$), for null circular geodesics. Indeed, the time
coordinate increases from $0\leq\alpha <\arcsin\sqrt{2/3}$
($0<r<R\sqrt{2/3}$), and decreases from
$\arcsin\sqrt{2/3}\leq\alpha <1$ ($R\sqrt{2/3}<r<R$). This is
plotted in Fig. \ref{Fig:circ}.

\subsubsection{Timelike geodesics}

For the case of an observer at rest with respect to the spacetime
geometry, i.e., with $(\alpha, \theta, \phi)$ fixed, we verify
that the relationship between the coordinate time and the proper
time measured by the observer is given by
\begin{equation}
t=\pm \sqrt{\cos\alpha}\;\tau  \,,
   \label{staticobserv}
\end{equation}
where the constant of integration has been defined as $t=0$ for
$\tau=0$.

Consider the specific case of $\theta=\pi/2$ and $d\phi=0$, which
implies $L=0$. Thus, we have to solve the following differential
equation
\begin{equation}
\frac{d\tau}{d\alpha}=\pm R({\cal E}^2\cos\alpha-1)^{-1/2} \,.
\end{equation}
Note the restriction ${\cal E}^2>1/\cos\alpha$. Integrating the
latter differential equation, we obtain
\begin{equation}
      \tau(\alpha) = \frac{\pm R}{\sqrt{{\cal E}^2-1}} {\rm F}(\alpha/2,2{\cal E}^2/({\cal E}^2-1))
      + C_2  \,,
\end{equation}
where ${\rm F}(\alpha,m)$ is the elliptic function of the fist
kind
\begin{equation}
      {\rm F}(\alpha,m)=\int_0^\alpha\,\frac{1}{\sqrt{1 - m \sin\negmedspace^2\varphi}}
      \,d\varphi \,.
\end{equation}

For the case of $\alpha={\rm const}$, we have
\begin{equation}
-1=-\frac{\dot{t}^2}{\cos\alpha}+R^2\sin^2\alpha\;\dot{\phi}^2 \,.
\end{equation}
Using the relationship $L=R^2\sin^2\alpha\,\dot{\phi}$, the latter
provides the following solution
\begin{equation}
t=\pm\sqrt{\cos\alpha\,\left(1+\frac{L^2}{R^2\sin^2\alpha}\right)}\;\;\tau
\,,
\end{equation}
which reduces to Eq.~(\ref{staticobserv}) if $L=0$. Note that one
may also find a relationship for the proper time measured by an
observer traversing a circumference at $\alpha={\rm const}$, in
terms of ${\cal E}$, given by
\begin{equation}
\tau=\frac{2\pi R\sin\alpha}{\sqrt{{\cal E}^2\cos\alpha-1}} \,.
\end{equation}

\section{Summary and Discussion}
\label{Sec:4}

In this work we took a fresh look at the Florides solution, which
represents an interior static and spherically symmetric perfect
fluid spacetime with vanishing radial stresses. In the standard
approach to its physical interpretation, the Schwarzschild vacuum
spacetime is matched at some constant radius hypersurface.
However, we were interested in the complete geometry of the matter
and therefore analyzed the geometry throughout the permitted range
of the radial coordinate without requiring the matching to an
exterior Schwarzschild spacetime. The resulting geometry is
particularly interesting since it admits a two-sphere singularity
which itself is the equator of a higher dimensional three-sphere.
This is quite contrary to the usual scenario where the
singularities are point-like.

The constant density Florides solution has an elegant interpretation as
the field inside an Einstein cluster which is generated by particles moving
in concentric circular orbits around the center. In view of this picture
the singular two-sphere can be interpreted as the surface where all the
particles are moving with the speed of light, and consequently our spacetime
picture breaks down as particles `behind' the singularity would move faster
than the speed of light.

In conclusion, we emphasize that spacetime singularities have
played a fundamental role in conceptual discussions of general
relativity. A key aspect of singularities in general relativity is
whether they are a disaster for the theory, as they imply the
breakdown of predictability. In this work, we have adopted the
attitude that the existence of singularities may be viewed as a
source to probe the foundations and limitations of general
relativity, and from which one may derive a valuable understanding
of gravitation, and in this context analyzed a particularly
interesting new type of a two-sphere singularity.

\acknowledgments We thank Roy Maartens for a careful reading of
the paper, for his valuable comments, and for bringing Ref.~\cite{Ellis}
to our attention. The work of CGB was supported by
research grant BO 2530/1-1 of the German Research Foundation
(DFG). FSNL was funded by Funda\c{c}\~{a}o para a Ci\^{e}ncia e
Tecnologia (FCT)--Portugal through the research grant
SFRH/BPD/26269/2006.

\end{document}

%% file: threesphere.pstex_t
\begin{picture}(0,0)%
\includegraphics{threesphere}%
\end{picture}%
\setlength{\unitlength}{4972sp}%
\begingroup\makeatletter\ifx\SetFigFont\undefined%
\gdef\SetFigFont#1#2#3#4#5{%
 \reset@font\fontsize{#1}{#2pt}%
 \fontfamily{#3}\fontseries{#4}\fontshape{#5}%
 \selectfont}%
\fi\endgroup%
\begin{picture}(2123,3364)(406,-2990)
\put(2026,-2941){\makebox(0,0)[lb]{\smash{{\SetFigFont{12}{14.4}{\familydefault}{\mddefault}{\updefault}{\color[rgb]{0,0,0}$\alpha=\pi/2$}%
}}}}
\put(901,-2806){\makebox(0,0)[lb]{\smash{{\SetFigFont{12}{14.4}{\familydefault}{\mddefault}{\updefault}{\color[rgb]{0,0,0}singular $\mathbb{S}^2$}%
}}}}
\put(811,-241){\makebox(0,0)[lb]{\smash{{\SetFigFont{12}{14.4}{\familydefault}{\mddefault}{\updefault}{\color[rgb]{0,0,0}$\mathbb{S}^3$}%
}}}}
\put(1756,254){\makebox(0,0)[lb]{\smash{{\SetFigFont{12}{14.4}{\familydefault}{\mddefault}{\updefault}{\color[rgb]{0,0,0}$\mathbb{S}^2$}%
}}}}
\put(406,-1456){\makebox(0,0)[lb]{\smash{{\SetFigFont{12}{14.4}{\familydefault}{\mddefault}{\updefault}{\color[rgb]{0,0,0}$\alpha=0$}%
}}}}
\end{picture}%

%% file: threesphere2.pstex_t
\begin{picture}(0,0)%
\includegraphics{threesphere2}%
\end{picture}%
\setlength{\unitlength}{4972sp}%
\begingroup\makeatletter\ifx\SetFigFont\undefined%
\gdef\SetFigFont#1#2#3#4#5{%
  \reset@font\fontsize{#1}{#2pt}%
  \fontfamily{#3}\fontseries{#4}\fontshape{#5}%
  \selectfont}%
\fi\endgroup%
\begin{picture}(4176,3397)(165,-3305)
\put(2026,-16){\makebox(0,0)[lb]{\smash{{\SetFigFont{12}{14.4}{\familydefault}{\mddefault}{\updefault}{\color[rgb]{0,0,0}identify}%
}}}}
\put(3376,-3256){\makebox(0,0)[lb]{\smash{{\SetFigFont{12}{14.4}{\familydefault}{\mddefault}{\updefault}{\color[rgb]{0,0,0}$\alpha=\pi/2$}%
}}}}
\put(676,-3256){\makebox(0,0)[lb]{\smash{{\SetFigFont{12}{14.4}{\familydefault}{\mddefault}{\updefault}{\color[rgb]{0,0,0}$\alpha=-\pi/2$}%
}}}}
\put(2116,-601){\makebox(0,0)[lb]{\smash{{\SetFigFont{12}{14.4}{\familydefault}{\mddefault}{\updefault}{\color[rgb]{0,0,0}$\mathbb{S}^3$}%
}}}}
\put(2026,-2581){\makebox(0,0)[lb]{\smash{{\SetFigFont{12}{14.4}{\familydefault}{\mddefault}{\updefault}{\color[rgb]{0,0,0}$\alpha=0$}%
}}}}
\end{picture}%